\documentclass[10pt,conference]{IEEEtran}
\IEEEoverridecommandlockouts 
\usepackage{blkarray}                                      
\usepackage{graphicx}                                      
\usepackage{amsmath}
\usepackage{amssymb}
\usepackage{amsfonts}
\usepackage{amsthm}
\usepackage[mathcal]{eucal}
\usepackage{mathrsfs}
\usepackage{booktabs}
\usepackage{enumerate}
\usepackage{multirow}
\usepackage[subrefformat=parens,farskip=0pt,justification=centering]{subfig}
\usepackage{color}
\usepackage{cite}                                          
\usepackage{comment}                                       
\usepackage{soul}                                          
\soulregister\cite7
\soulregister\ref7
\soulregister\pageref7
\usepackage{etoolbox}                                      
\usepackage{url}
\usepackage{nth}                                           
\usepackage{bm}                                            
\usepackage{courier}
\usepackage{balance}
\usepackage{threeparttable}
\usepackage{xcolor,colortbl}
\usepackage{footnote}
\usepackage{listings}
\usepackage{setspace}                                      
\usepackage[inline]{enumitem}
\usepackage{verbatim}
\usepackage[bookmarks=false]{hyperref}
\hypersetup{
    colorlinks = true,
    citecolor  = blue,
    linkcolor  = blue,
    urlcolor   = blue,
}
\usepackage{tikz}
\usetikzlibrary{patterns,snakes}
\usetikzlibrary{positioning,calc,fit,decorations.pathmorphing,shapes.geometric, shapes.gates.logic.US, calc}
\usetikzlibrary{arrows,arrows.meta,decorations.markings,shapes,shapes.arrows}
\usetikzlibrary{decorations,decorations.pathreplacing}
\usetikzlibrary{backgrounds}
\usepackage{filecontents}                                  
\usepackage{pgfplots}
\usepackage{pgfplotstable}
\usepgfplotslibrary{groupplots}
\usepackage{scalefnt}
\pgfplotsset{compat=newest}
\usepackage{caption}
\usepackage{pifont}                                        
\usepackage{cleveref}
\crefname{algocf}{algorithm}{algorithms}
\Crefformat{figure}{Fig.~#2#1#3}                           
\Crefname{subfigure}{Fig.}{Figs.}
\Crefname{figure}{Fig.}{Figs.}
\Crefformat{table}{TABLE~#2#1#3}                           
\usepackage[figuresright]{rotating}

\iffalse
\usepackage{algorithm}
\usepackage{algpseudocode}                                 
\algrenewcommand\textproc{\texttt}
\makeatletter
\let\OldStatex\Statex
\renewcommand{\Statex}[1][3]{%
  \setlength\@tempdima{\algorithmicindent}%
  \OldStatex\hskip\dimexpr#1\@tempdima\relax
}
\makeatother
\else
\usepackage[linesnumbered,ruled,vlined]{algorithm2e}
\fi

\definecolor{CUHKorange}{RGB}{244,106,18} 
\definecolor{CUHKblue}{RGB}{0,111,190}    
\definecolor{CUHKgreen}{RGB}{0,127,128}   
\definecolor{CUHKred}{RGB}{228,46,36}     
\definecolor{CUHKyellow}{RGB}{198,148,34} 
\definecolor{CUHKdark}{RGB}{114,44,114}   
\definecolor{CUHKmiddle}{RGB}{144,44,144} 
\definecolor{CUHKlight}{RGB}{167,44,167} 
\definecolor{CUHKpurple}{RGB}{117,15,109}
\definecolor{CUHKgold}{RGB}{221,163,0}
\definecolor{CUHKribbon}{RGB}{244,223,176}
\definecolor{CUHKblack}{RGB}{34,24,21}

\renewcommand{\bf}[1]{\textbf{#1}}
\renewcommand{\tt}[1]{\texttt{#1}}
\renewcommand{\vec}[1]{\boldsymbol{#1}}    

\newcommand{\minisection}[1]{\vspace{.02in}\noindent{\textbf{#1}}.}

\usepackage{tcolorbox}
\tcbuselibrary{skins,breakable}
    {\endtcolorbox}
    {\endtcolorbox}

\usepackage[top=0.80in,bottom=0.88in,left=0.62in,right=0.62in]{geometry}
\iftrue
\newcommand{\subparagraph}{}
\usepackage{titlesec}
\titlespacing*{\section}{0pt}{1.8ex plus .2ex minus .2ex}{0.4ex plus .2ex}
\titlespacing*{\subsection}{0pt}{1.0ex plus .2ex minus .2ex}{0.2ex plus .2ex}
\fi

\crefname{mytheorem}{Theorem}{Theorems}
\crefname{mylemma}{Lemma}{Lemmas}
\crefname{myclaim}{Claim}{Claims}
\crefname{myproperty}{Property}{Properties}
\crefname{mycorollary}{Corollary}{Corollaries}

\RequirePackage[normalem]{ulem} 
\RequirePackage{color}\definecolor{RED}{rgb}{1,0,0}\definecolor{BLUE}{rgb}{0,0,1} 

\usepackage{times}
\usepackage{listings}
\usepackage{balance}
\graphicspath{{./}}

\newif\ifrev
\revtrue
\ifrev
  \newcommand{\bei}[1]{{\color{red} [Bei: #1]}} 
\else
  \newcommand{\bei}[1]{}
\fi

\definecolor{myorange}{RGB}{238,97,42}  %
\definecolor{myblue}{RGB}{178,179,249}  
\definecolor{mygrey}{RGB}{166,166,166}  %
\definecolor{mygreen}{RGB}{180,210,36}  
\definecolor{myred}{RGB}{238,0,0}       
\definecolor{myyellow}{RGB}{198,148,34} 
\definecolor{mydark}{RGB}{114,44,114}   
\definecolor{mymiddle}{RGB}{144,44,144} 
\definecolor{mylight}{RGB}{167,44,167}  
\definecolor{myblue1}{RGB}{137,157,192}  
\definecolor{mygreen1}{RGB}{69,137,148}  

\begin{document}

\title{
LongRTL: Graph-Similarity-Guided LLM-driven Long Context RTL Optimization
}

\author{
    Yuyang Ye$^{1}$, 
    Che-Kuan Shen$^{2}$,
    Xiangfei Hu$^{3}$,
    Yuchen Liu$^{3}$,
    Shuo Yin$^{1}$,
    Xufeng Yao$^{1}$,
    Bei Yu$^{1, \dag}$,
    Tsung-Yi Ho$^{1}$\\
    \thanks{
        $^{\dag}$Corresponding author.
    }
    \thanks{
        This work is partially supported by
        The Research Grants Council of Hong Kong SAR (No.~CUHK14211824 and No.~CUHK14210723).
    }
    $^1$CUHK \quad
    $^2$National Central University \quad
    $^3$Southeast University \\
}

\maketitle
\pagestyle{empty}

\begin{abstract}
Large Language Models (LLMs) show great promise in RTL code generation and optimization. 
However, real-world RTL designs are typically long, entangled, and poorly modularized—posing a major challenge due to context-length limitations and lack of structure. 
To overcome these obstacles, we propose a scalable LLM-based RTL optimization framework guided by graph similarity. 
Our method introduces three collaborative agents:
(1) a Partition Agent that decomposes RTL designs into semantically meaningful AST subtrees, guided by AST graph similarity to reusable design templates;
(2) an Optimization Agent that generates RTL submodule code based on partitioned subtrees using multi-modal Retrieval-Augmented Generation (RAG) with both AST and RTL guidance; and
(3) a Reconstruction Agent that reassembles optimized submodules based on logic-aware ordering and Graph-RAG prompting, ensuring global functional equivalence.
Together, these components enable robust, structure-aware optimization of long-context RTL designs, bridging the gap between toy examples and industrial-scale hardware codebases.
\end{abstract}

\section{Introduction}
\label{sec:intro}
Large Language Models (LLMs) have recently demonstrated strong capabilities in hardware design tasks such as RTL code generation \cite{rtlcoder,thakur2024verigen,zehua2024betterv}, optimization \cite{rtlrewriter,wang2025symrtlo}, and verification \cite{hu2024uvllm,wang2024chatcpu}.  
However, existing LLM-based RTL optimization approaches are largely constrained to short, modular code snippets that neatly fit within the context window of modern LLMs.  
In contrast, real-world industrial RTL designs are often long, monolithic, and lack clear module boundaries, with complex interdependencies spanning thousands of lines.  
This disparity fundamentally limits the scalability and practicality of current LLM-driven techniques.

A key challenge lies in the context-length bottleneck: LLMs struggle to process global design semantics and long-range dependencies in oversized RTL designs \cite{scalertl}.  
Effective LLM-based optimization at scale thus requires a principled workflow to:  
(1) partition the design into meaningful and optimizable units,  
(2) perform structure- and function-aware optimization for each part, and  
(3) reconstruct the design while preserving overall functional equivalence.

\begin{figure}[tb!]
    \centering
    \includegraphics[width=.98\linewidth]{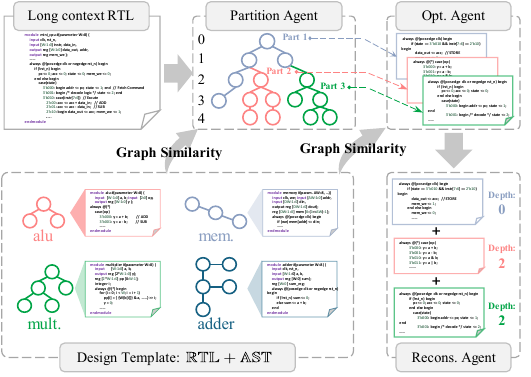}
    \caption{Overall flow of our work.}
    \label{fig:flow}
\end{figure}
To address this, we propose a graph similarity–guided LLM optimization framework for long-context RTL designs.  
Inspired by how human engineers recognize and reuse known design idioms, our method employs AST-level graph similarity to decompose, optimize, and reconstruct RTL  in a modular and semantically consistent manner.  
As shown in \Cref{fig:flow}, our framework comprises three collaborative agents:

\noindent \textbf{Partition Agent.}  
We parse RTL into abstract syntax trees (ASTs) and reason at the behavioral level.
A library of high-quality design templates is built, each paired with its RTL text and AST graph.
Given a long RTL input, we compute AST-level graph similarity using a trained GNN, and apply a tree-based dynamic programming algorithm to partition the global AST into $K$ connected subtrees that best match known templates.
These subtrees represent reusable design patterns and enable interpretable decomposition —mimicking how engineers reason about large RTL designs.

\noindent \textbf{Optimization Agent.}  
Each partitioned subtree is mapped back to RTL and optimized by an LLM agent equipped with a multi-modal Retrieval-Augmented Generation (RAG) mechanism.
Instead of relying solely on RTL text, our framework jointly leverages both RTL code and AST graph structure during retrieval.  
For each subgraph, we retrieve a small set of RTL+AST template exemplars with similar functionality and structural hierarchy.  
This targeted retrieval narrows the search space and provides directionally accurate optimization.
We further explore multiple rewrites using Monte Carlo Tree Search (MCTS), and select optimized RTL submodules with superior design performance metrics (e.g., delay, area, power).

\noindent \textbf{Reconstruction Agent.}  
Optimized submodules are reassembled into a functionally equivalent design using a logic-aware hierarchical strategy.
Crucially, we introduce a graph RAG prompt mechanism that encodes the structural relationship between each partitioned subtree and the original full AST.
This enables the LLM to reason about interconnections and avoid logic or signal mismatches.
This reconstruction respects logic depth and control hierarchy, ensuring function equivalence and smooth end-to-end synthesis.

Overall, we present the first end-to-end framework that enables practical LLM-based RTL optimization for long-context RTL designs.  
By combining AST graph partition, targeted multimodal retrieval, and logic-guided reconstruction, our method significantly extends the capability of LLMs beyond toy examples —paving the way for real-world deployment in RTL code optimization.  
Our contributions are summarized as follows:
\begin{itemize}
    \item We propose a graph similarity–guided RTL partitioning approach that segments large-scale designs into semantically meaningful subtrees.
    \item We develop a multi-modal RAG system that combines RTL text and AST graph as multimodal guidance, enabling structurally grounded and functionally equivalent RTL submodule generation and optimization based on partitioned subtrees.
    \item We introduce a logic-aware reconstruction strategy that integrates optimized submodules into the RTL design while preserving function equivalence, using a graph-based prompt to connect local and global structure.
    \item We validate our framework on long-context benchmarks, achieving 100\% functional equivalence and significant PPA improvements.
\end{itemize}

\section{Preliminaries}
\label{sec:pre}

\minisection{Abstract Syntax Tree (AST)}
RTL designs can be naturally represented as Abstract Syntax Trees (ASTs), where each node denotes a code construct (e.g., module, always block, if-statement), and edges reflect parent-child syntax relationships.  
Compared to netlists or control/data-flow graphs, ASTs preserve syntactic structure and semantic hierarchy, making them more suitable for RTL code understanding and transformation—especially when paired with LLMs.

\minisection{Graph Similarity}
Graph similarity measures how structurally and semantically aligned two graphs are.  
Given two AST graphs $\mathbb{G}_1$ and $\mathbb{G}_2$, a similarity score $S(\mathbb{G}_1, \mathbb{G}_2) \in [0,1]$ is computed based on topological matching, label consistency, and local subgraph overlap.  
In our work, graph similarity enables us to match AST subtrees of long context RTL with known functional patterns (e.g., ALUs, FSMs).

\section{Method}
\label{sec:method}
\subsection{Design Templates}
\label{sec:data}
To support graph-guided partitioning, retrieval-based optimization, and logic-aware reconstruction, we construct a reusable design template at both RTL and AST levels.

\noindent \textbf{RTL Design Templates:}  
We collect a diverse set of behavioral RTL designs covering common circuit functions: arithmetic units (adder, subtractor, multiplier), data operations (left/right shifter, barrel shifter, bitwise logic), control structures (multiplexer, comparator, priority encoder, decoder), and storage elements (latch, register, counter). 
Each RTL template is written in synthesizable Verilog and encapsulates a reusable functional unit.

\noindent \textbf{AST Graph Templates:}  
Each RTL template is parsed into an Abstract Syntax Tree (AST), encoded as a graph $\mathbb{G} = (\vec{A}, \vec{F})$. 
Here, $\vec{A}$ denotes the adjacency matrix encoding the parent–child structure of the AST, and $\vec{F}$ is the node feature matrix enables grap similarity computation.
Nodes in $\mathbb{G}$ are categorized as:
(1) Operation nodes: e.g., \texttt{Assign}, \texttt{BinOp}, \texttt{If}, \texttt{Cond}, \texttt{Always}, etc., capturing control and logic structures;
(2) Port nodes: e.g., \texttt{input}, \texttt{output}, \texttt{wire}, \texttt{reg}, \texttt{inout}, representing interface and connectivity.

Each node is annotated with a one-hot encoded feature vector $\vec{f}_v \in \vec{F}$ based on its type.
We develop an AST-based template library $\mathcal{G} = \{\mathbb{G}_1, ..., \mathbb{G}_N\}$, extracted from RTL design exemplars, to support downstream tasks.

\begin{figure}[tb!]
    \centering
    \includegraphics[width=.99\linewidth]{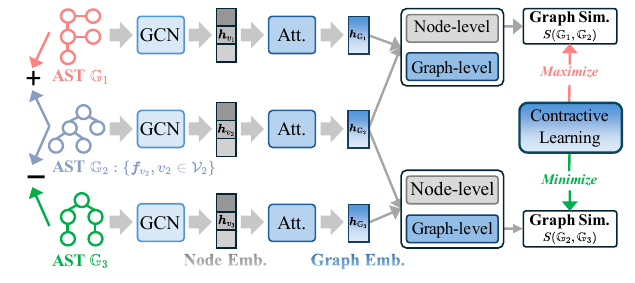}
    \caption{Training of AST graph similarity computation.}
    \label{fig:sim}
\end{figure}
\subsection{AST Graph Similarity Computation}
\label{sec:similarity}
As illustrated in \Cref{fig:sim}, we measure the graph similarity $S(\mathbb{G}_i, \mathbb{G}_j)$ between two AST graphs $\mathbb{G}_i$ and $\mathbb{G}_j$ via three stages:  (1) node-level similarity, (2) graph-level similarity, and (3) contrastive training.

\noindent \textbf{Node-level similarity.}
In AST graphs, each AST node feature $\vec{f}_v$ is encoded using a Graph Convolutional Network (GCN) to propagate local structural information and generate node encoding results $\vec{h}_v$:
\begin{equation}
\vec{h}_v = \mathrm{GCN}(\vec{f}_v).
\end{equation}

To compare ASTs at a fine-grained level, we define their node-level similarity as the average cosine similarity over all node pairs based on node encoding results:
\begin{equation}
\label{eq:sn}
S_{\text{node}}(\mathbb{G}_i, \mathbb{G}_j) = \frac{1}{|\mathcal{V}_i| \cdot |\mathcal{V}_j|} \sum_{u \in \mathcal{V}_i} \sum_{v \in \mathcal{V}_j} \cos(\vec{h}_u, \vec{h}_v),
\end{equation}
where $\mathcal{V}_i$ and $\mathcal{V}_j$ are node sets of AST graphs $\mathbb{G}_i$ and $\mathbb{G}_j$.
$h_u$ and $h_v$ are the final node encoding results in $G_i$ and $G_j$ after graph learning, respectively.
$\cos(h_u, h_v)$ denotes the cosine similarity between the two node vectors.

\noindent \textbf{Graph-level similarity.}
To capture global semantics, we apply attention-based pooling over node encoding results to generate graph embeddings $\vec{h}_\mathbb{G}$:
\begin{equation} 
\alpha_v = \mathrm{sigmoid}(\vec{w}^\top \vec{h}_v), \quad \vec{h}_\mathbb{G} = \sum_{v \in \mathcal{V}} \alpha_v \vec{h}_v,
\end{equation}
where $w$ is a learnable vector. The graph-level similarity between two AST graphs is computed via:
\begin{equation}
\label{eq:sg}
S_{\text{graph}}(\mathbb{G}_i, \mathbb{G}_j) = \cos(h_{\mathbb{G}_i}, h_{\mathbb{G}_j}),
\end{equation}
$\cos(h_{\mathbb{G}_i}, h_{\mathbb{G}_j})$ denotes the cosine similarity between two graph embeddings.
Then, the graph similarity of $\mathbb{G}_i$ and $\mathbb{G}_j$ is computed via:
\begin{equation}
\label{eq:s}
S(\mathbb{G}_i, \mathbb{G}_j) = S_{\text{node}}(\mathbb{G}_i, \mathbb{G}_j) +S_{\text{graph}}(\mathbb{G}_i, \mathbb{G}_j).
\end{equation}

\noindent \textbf{Contrastive learning.}
To enforce functional consistency across structurally diverse ASTs, we adopt contrastive learning over the graph learning training.
We define a positive pair $(\mathbb{G}_i, \mathbb{G}_j^+)$ if both ASTs correspond to the same design functionality,
and define negative pairs $(\mathbb{G}_i, \mathbb{G}_j^-)$ if they belong to different RTL design types.

Based on node-level and graph-level graph similarity, we optimize the following InfoNCE-style \cite{oord2018info} contrastive loss:
\begin{equation}
\mathcal{L}_{\text{CL}} = -\log \frac{ \exp\left( S(\mathbb{G}_i, \mathbb{G}_j^+) / \tau \right) }{ \exp\left( S(\mathbb{G}_i, \mathbb{G}_j^+) / \tau \right) + \sum^{\mathcal{N}_{-}} \exp\left(S(\mathbb{G}_i, \mathbb{G}_j^-) / \tau \right) }, 
\end{equation}
where $S(\mathbb{G}_i, \mathbb{G}_j)$ is the graph similarity between two graphs computed via \Cref{eq:s}.
$\mathcal{N}_{-}$ is the number of negative samples.
$\tau$ is a temperature hyperparameter (set to 0.1).

This training objective pulls together functionally equivalent AST graphs and pushes apart those with different semantics.  
Through this contrastive supervision, the learned AST embeddings reflect both structural features and functional alignment.  
It enhances the reliability of similarity-based matching in downstream partitioning and optimization.

\subsection{Partition Agent}
\label{sec:p}
To support scalable optimization of long-context RTL designs, we propose a Tree-based Dynamic Programming (Tree-DP) algorithm that hierarchically partitions a large AST graph $\mathbb{G}_{\text{long}}$ into $K$ disjoint, connected subtrees $\{\mathbb{G}_{s_1}, \dots, \mathbb{G}_{s_K}\}$.
Tree-DP performs graph similarity-guided partitioning, where each subtree is matched against the reusable design template library $\mathcal{G}$ (described in \Cref{sec:data}) to ensure functional alignment and structural coherence.
It ensures that the resulting subtrees resemble known RTL modules (e.g., adders, shifters, FSMs).  
This encourages each subtree to exhibit well-formed functional intent, making it more interpretable for LLMs and easier to optimize, regenerate, or replace in downstream tasks.
After partitioning, each subtree serves as an LLM-optimizable unit and is passed to the downstream optimization agent.
The complete procedure is detailed in \Cref{alg:1}.

\noindent\textbf{Objective.}
The partitioning process is guided by an AST graph similarity function $S(\cdot, \cdot)$ described in \Cref{sec:similarity}. 
For each candidate subtree $\mathbb{G}_{s_k}$ extracted from $\mathbb{G}_{\text{long}}$, we identify its most similar design pattern from the template library $\mathcal{G}$, and define the template relevance score as the maximum graph similarity.
Our global objective is thus to find a partition configuration that maximizes total similarity: \begin{equation}
\max_{\{\mathbb{G}_{s_1}, \dots, \mathbb{G}_{s_K}\}} \sum_{k=1}^{K} \max_{\mathbb{G}_{\text{temp}} \in \mathcal{G}} S(\mathbb{G}_{s_k}, \mathbb{G}_{\text{temp}}). 
\end{equation}

\noindent\textbf{Constraints.}  
To ensure both validity and downstream usability, we impose the following constraints on the resulting partitions: 
(i) Each subtree $\mathbb{G}_{s_k}$ must be connected and correspond to a syntactically and semantically valid RTL fragment; 
(ii) The union of all subtrees must exactly cover the full AST: $\bigcup_{k=1}^K \mathbb{G}_{s_k} = \mathbb{G}_{\text{long}}$, with $\mathbb{G}_{s_i} \cap \mathbb{G}_{s_j} = \emptyset$ for all $i \ne j$; 
(iii) The number of partitions $K$ is either user-specified or adaptively determined based on the design scale and optimization budget.

\noindent\textbf{Workflow.} 
Exploiting the hierarchical structure of ASTs, Tree-DP performs partitioning via four key stages:
(1) Traversal: 
A bottom-up traversal is performed over the AST $\mathbb{G}_{\text{long}}$, visiting each node $v$ and recursively computing the best partition scores for subtrees rooted at $v$ under varying partition budgets.
(2) Merge \& Score:  
At each internal node, we enumerate valid combinations of child subtrees and attempt to merge them into higher-level candidates. 
Each merged subtree $\mathbb{G}_{s_k}$ is compared against the design template library $\mathcal{G}$ based on graph similarity.
(3) Update:
Among all merge candidates, only the one with the highest cumulative similarity is preserved in the DP table for each partition size $k$ during the bottom-up traversal from leaves to node $v$
(4) Backtrack:
After the DP table is populated, Tree-DP traces back from the root of $\mathbb{G}_{\text{long}}$ to recover the partition plan $\{\mathbb{G}_{s_1}, \dots, \mathbb{G}_{s_K}\}$, ensuring that the selected subtrees maximize similarity to known design patterns in $\mathcal{G}$.

\begin{algorithm}[tb!]
\caption{Tree-DP Partitioning for Large AST}
\footnotesize
\label{alg:1}
\KwIn{Long RTL AST graph $\mathbb{G}_{\text{long}}$, template library $\mathcal{G}$, target partition count $K$}
\KwOut{Partitioned subtrees $\{\mathbb{G}_{s_1}, \dots, \mathbb{G}_{s_K}\}$}

\BlankLine
\textbf{function} \textsc{BestMatchTemplate}($\mathbb{G}_{\text{cand}}, \mathcal{G}$):\\
\quad \Return $\max\limits_{\mathbb{G}_{\text{temp}} \in \mathcal{G}} S(\mathbb{G}_{\text{cand}}, \mathbb{G}_{\text{temp}})$\;
\tcp*{Graph similarity-driven searching}

\BlankLine
\textbf{function} \textsc{TreeDP}($v$): 
\ForEach{child $c$ of $v$}{\underline{\texttt{Traverse}} from leaves to root \textsc{TreeDP}($c$)}
\For{$k = 1$ \KwTo $K$}{
  \ForEach{valid $k$-way partition of $\text{children}(v)$}{
    \underline{\texttt{Merge}} children into $\mathbb{G}_{\text{cand}}$\;
    $s \leftarrow \textsc{BestMatchTemplate}(\mathbb{G}_{\text{cand}}, \mathcal{G})$\;
    $S_{total} \leftarrow s + \sum_{j=1}^{k} \text{DP}[c_j][k_j]$\;
    \If{$S_{total}$ $>$ $\text{DP}[v][k]$}{
      \underline{\texttt{Update}} best config $\text{DP}[v][k] \leftarrow S_{total}$; \;
    }
  }
}
\Return $\text{DP}[v]$

\BlankLine
\textbf{Main execution:}\\
Initialize $\text{DP}[v][k] = -\infty$ for all $v, k$\;
\textsc{TreeDP}(root of $\mathbb{G}_{\text{long}}$)\;
\underline{\texttt{Backtrack}} from $\text{DP}[\text{root}][K]$ to get $\{\mathbb{G}_{s_1}, \dots, \mathbb{G}_{s_K}\}$\;
\end{algorithm}

\subsection{Optimization Agent}
\label{sec:optimization}
\begin{figure}[tb!]
    \centering
    \includegraphics[width=.95\linewidth]{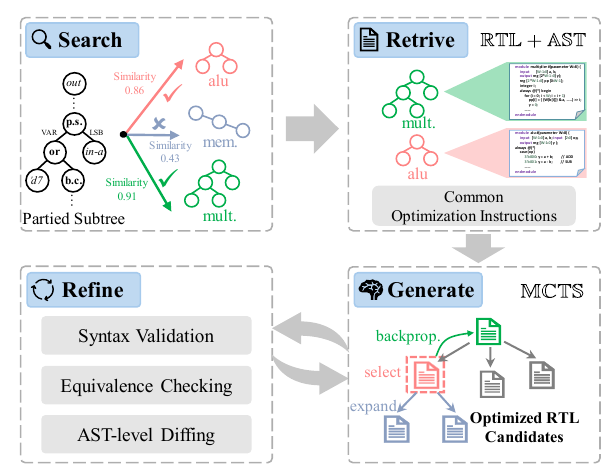}
    \caption{The optimization agent pipeline.}
    \label{fig:opt}
\end{figure}

To enable structure-preserving and high-quality RTL optimization of partitioned AST subtrees, we develop an optimization agent based on AST+RTL multimodal retrieval-augmented large language models (LLMs).  
This agent follows a structured four-stage pipeline: Search, Retrieve, Generate, and Refine, as illustrated in \Cref{fig:opt}.  
For each AST subtree $\mathbb{G}_{s_i}$ obtained from the partition agent, the optimization agent searches for AST structurally aligned RTL exemplars, explores diverse rewrite candidates through Monte Carlo Tree Search (MCTS), and iteratively refines outputs based on functional equivalence feedback.
Our method achieves faithful optimization of RTL submodules, supporting scalable long-context RTL design optimization.

\noindent\textbf{Search.}  
Given a partitioned AST subgraph $\mathbb{G}_{s_i}$, we first encode it into a graph embedding using a  trained GNN for computing AST-level similarity (\Cref{sec:similarity}).  
We then perform a graph similarity search over the AST template database to retrieve the top-$N$ structurally similar AST designs.  
This step constructs a high-quality candidate set that reflects similar control and computation patterns, forming a strong foundation for downstream RTL generation.

\noindent\textbf{Retrieve.}  
Each retrieved AST $\mathbb{G}_j$ is paired with its corresponding RTL implementation $\text{RTL}_j$ to form a multi-modal exemplar set $\{(\mathbb{G}_j, \text{RTL}_j)\}$.  
These exemplars—comprising both structural and semantic information—are used to prompt the LLM in a retrieval-augmented generation (RAG) setting.  
Unlike static prompt engineering, our retrieval dynamically adapts to the structure of $\mathbb{G}_{s_i}$, enabling query-specific generation that respects the design intent.

\noindent\textbf{Generate.}  
Using the exemplar pairs as prompt, the LLM generates candidate rewrites $\text{RTL}_{s_i}'$ for the target partitioned subtree $\mathbb{G}_{s_i}$.  
To systematically explore diverse rewrites, we employ Monte Carlo Tree Search (MCTS), where each node in the search tree corresponds to a generated RTL variant.  
Each expansion samples different prompt configurations (e.g., reordering exemplars), and each candidate is evaluated using a hybrid score:
\begin{equation}
\text{Score}(\text{RTL}_i') = \lambda \cdot S(\mathbb{G}_i, \mathbb{G}_i') + (1 - \lambda) \cdot \text{PPA}( \text{RTL}_{s_i}'),
\end{equation}
where $S(\mathbb{G}_i, \mathbb{G}_i')$ measures the graph-level structural similarity between the original and generated ASTs,  
and $\text{PPA}(\text{RTL}_{s_i}')$ denotes a performance-aware proxy metric reflecting the RTL’s quality.  
The PPA metric is defined as the normalized average of delay, area, and power of generated RTL.  
It can be reweighted by the designer to prioritize specific objectives, such as low power or high speed.

\noindent\textbf{Refine.}  
Once MCTS completes—either by rollout limit or convergence, we select the top-scoring candidate as the optimized RTL.  
The selected RTL undergoes a multi-stage verification pipeline:  
(1) Syntax validation via Verilog compiler checks;  
(2) Functional simulation under benchmark inputs to ensure functional equivalence;  
(3) AST-level diffing to capture unintended logic drift.  
If any issues are detected, the agent reuses the MCTS procedure with refined prompts until a valid and performant RTL variant is found.


\subsection{Reconstruction Agent}
\label{sec:reconstruction} 
\begin{figure}[tb!]
    \centering
    \includegraphics[width=0.95\linewidth]{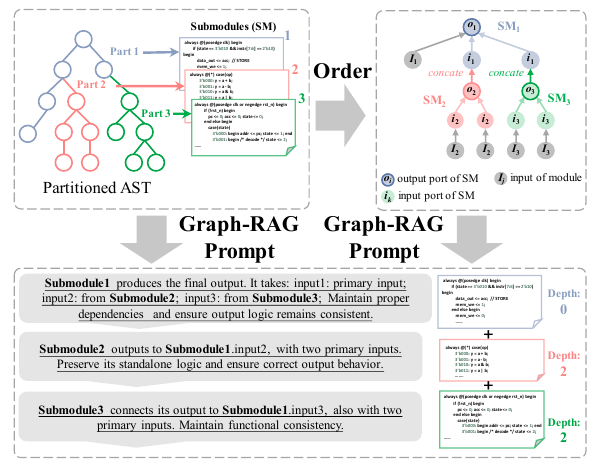}
    \caption{Graph-RAG prompting via logic-aware ordering.}
    \label{fig:rec}
\end{figure}

After modular optimization, the reconstruction agent reassembles optimized RTL submodules $\text{RTL}_{s_i}$ of partitioned AST subtrees into a functional equivalent design.  
To support this process, we employ a logic-aware ordering strategy and a Graph-RAG prompting framework, enabling the LLM to generate the functional equivalent top-level RTL with accurate connectivity and control flow, as shown in \Cref{fig:rec}.

\noindent\textbf{Order.}  
Each optimized RTL submodule $\text{RTL}_{s_i}$ corresponds to a partitioned AST subtree $\mathbb{G}_{s_i}$.  
To preserve functional semantics and signal dependencies, we sort the submodules based on their logical depth and connectivity within the large AST $\mathbb{G}_{\text{long}}$ of the original long context RTL.  
This facilitates correct instantiation and signal propagation.

\noindent\textbf{Graph-RAG Prompting.}  
We construct a hierarchical Graph-RAG prompt that jointly encodes structural, semantic, and hierarchical information.  
Specifically, the prompt includes: (i) the set of partitioned AST subtrees $\mathbb{G}_{s_i}$ and their corresponding RTLs $\text{RTL}_{s_i}$; (ii) the full large AST $\mathbb{G}_{\text{long}}$ to retain global module hierarchy and I/O interconnection patterns; and (iii) the original long context RTL for semantic grounding.  
This fusion enables the LLM to reason over hierarchical control flow, structural alignment, and naming consistency, producing a \texttt{top.v}-style RTL that integrates all submodules coherently.

\noindent\textbf{Validate.}  
We apply the same validation pipeline used in the optimization agent.  
The generated RTL is checked for syntax errors using Verilog compilers, and functionally verified via equivalence checking.  
We also perform AST-level diffing to detect structural drift.  
If inconsistencies are found, the model is re-prompted with updated context to regenerate the erroneous segments.  
This iterative refinement ensures that the final top-level RTL design is functionally equivalent and structurally faithful to the original RTL design.

\section{Experimental Results}
To evaluate our proposed framework, we implement a complete RTL optimization pipeline. The main components are as follows:
For AST parsing and preprocessing, we use \texttt{PyVerilog} \cite{takamaeda2015pyverilog} to parse input RTL code into abstract syntax trees (ASTs).
For the partition agent, we implement a Tree-based Dynamic Programming (Tree-DP) algorithm in Python, supporting configurable partition budgets and similarity thresholds. Depending on design scale, the partition count $K$ is adaptively chosen from {5, 10, 15, 20}.
For the optimization and reconstruction agents, we build both upon \texttt{ChatGPT 4o} \cite{hurst2024gpt} via OpenAI API. 
The optimization follows the four-stage RAG-based workflow described in \Cref{sec:optimization}, while reconstruction leverages a graph-RAG flow with logic-depth–aware prompting, as in \Cref{sec:reconstruction}.
The graph similarity module uses a trained Graph Convolutional Network (GCN) \cite{kipf2016gcn} to guide both retrieval and partitioning.
For PPA Evaluation, all RTL designs are synthesized using \texttt{Synopsys Design Compiler} \cite{dc} under the \texttt{ASAP 7nm technology node} \cite{clark2016asap7}. We report post-synthesis delay, area, and power, and compute a normalized PPA score as a weighted average. The weights are user-configurable to support various optimization priorities (e.g., timing or power).
For verification, we adopt a multi-stage correctness check to ensure functional and structural equivalence. First, syntax is validated via \texttt{iVerilog} \cite{williams2002iverilog}. Then, we perform simulation-based functional checking with Python-generated testbenches. Additionally, we employ:
(i) \texttt{ABC} \cite{brayton2010abc} for SAT-based combinational equivalence checking, and
(ii) \texttt{egg} \cite{coward2022egg} for efficient arithmetic reasoning.
\begin{table}[t]
	\centering
	\caption{Benchmark characteristics of single-module long-context RTL designs. Delay ($ps$), Area ($\mu m^2$), and Power ($mW$) are reported by \texttt{Synopsys Design Compiler}.
}
	\label{tab:slong}
	\resizebox{.9\linewidth}{!}
    {
\begin{tabular}{|ccccccc|}
\hline
Des.          & Lines & Nodes & Wires & Delay & Area & Power \\ \hline \hline
add64           & 195       & 917   & 1306  & 47.6     & 173.9    & 39.9     \\
mult32          & 88        & 2119  & 2873  & 617.2    & 230.2    & 194.3    \\
comput    & 204       & 1520  & 2265  & 952.5     & 141.6    & 116.8    \\
traffic         & 90        & 66    & 69    & 9.6      & 59.7     & 1.4      \\
alu             & 111       & 909   & 1110  & 1555.6    & 79.1     & 49.3     \\
radix      & 300       & 300   & 439   & 269.7    & 29.9     & 6.7      \\
asyn      & 147       & 560   & 601   & 122.7    & 85.2     & 13.5     \\
accu            & 77        & 66    & 111   & 61.2     & 9.8      & 3      \\
fpu\_pre  & 235       & 663   & 881   & 40.3     & 65.0        & 46.9     \\
fpu\_post & 638       & 1692  & 2119  & 3009.4   & 138.5    & 102.5    \\
mc\_sel    & 201       & 380   & 464   & 299.1    & 49.9     & 28.6     \\
mc\_rf      & 166       & 193   & 256   & 126.9    & 32.8     & 6.6      \\
buffer     & 268       & 774   & 847   & 54.2     & 103.7    & 24.2     \\
eth      & 325       & 527   & 739   & 55.3      & 87.9     & 12.7      \\ \hline \hline
AVE.        & 217       & 763   & 1005  & 515.8       & 92.0        & 46.2        \\ \hline
\end{tabular}
}
\end{table}
\begin{table}[t]
	\centering
	\caption{Benchmark characteristics of multi-module long-context RTL designs.}
	\label{tab:mlong}
	\resizebox{.9\linewidth}{!}
    {
\begin{tabular}{|ccccccc|}
\hline
Des.          & Lines & Nodes & Wires & Delay & Area & Power \\ \hline \hline
ac97 & 2030  & 8214  & 8243  & 89.9  & 1173.2 & 30.2  \\
aes  & 2195  & 7043  & 7953  & 105.7 & 922.9  & 14.6  \\
crca & 1034  & 3812  & 3855  & 105.3 & 340.5  & 3.1   \\
ecg  & 1525  & 66872 & 67385 & 305.6 & 6636.4 & 41.1  \\ \hline \hline
AVE. & 1696  & 21485 & 21859 & 151.6 & 2268.3 & 22.2  \\ \hline
\end{tabular}
}
\end{table}

\begin{table}[t]
	\centering
	\caption{Functional equivalence pass rate results.}
	\label{tab:equ}
	\resizebox{0.95\linewidth}{!}
    {
\begin{tabular}{|c|c|c|c|c|c|c|}
\hline
Des. & Y+E & GPT4o & Gemini & Coder & Rewriter & Ours \\ \hline \hline
add64     & $\checkmark$ & $\times$ & $\checkmark$ & $\times$ & $\checkmark$ & \textbf{$\checkmark$} \\
mult32    & $\checkmark$ & $\checkmark$ & $\checkmark$ & $\checkmark$ & $\checkmark$ & \textbf{$\checkmark$} \\
comput    & $\checkmark$ & $\times$ & $\times$ & $\times$ & $\times$ & \textbf{$\checkmark$} \\
traffic   & $\checkmark$ & $\checkmark$ & $\checkmark$ & $\checkmark$ & $\checkmark$ & \textbf{$\checkmark$} \\
alu       & $\checkmark$ & $\checkmark$ & $\checkmark$ & $\checkmark$ & $\checkmark$ & \textbf{$\checkmark$} \\
radix     & $\checkmark$ & $\times$ & $\times$ & $\times$ & $\times$ & \textbf{$\checkmark$} \\
asyn      & $\checkmark$ & $\checkmark$ & $\times$ & $\checkmark$ & $\checkmark$ & \textbf{$\checkmark$} \\
accu      & $\checkmark$ & $\checkmark$ & $\checkmark$ & $\checkmark$ & $\checkmark$ & \textbf{$\checkmark$} \\
fpu\_pre  & $\checkmark$ & $\times$ & $\times$ & $\times$ & $\times$ & \textbf{$\checkmark$} \\
fpu\_post & $\checkmark$ & $\times$ & $\times$ & $\times$ & $\times$ & \textbf{$\checkmark$} \\
mc\_sel   & $\checkmark$ & $\times$ & $\times$ & $\times$ & $\times$ & \textbf{$\checkmark$} \\
mc\_rf    & $\checkmark$ & $\checkmark$ & $\times$ & $\checkmark$ & $\checkmark$ & \textbf{$\checkmark$} \\
buffer    & $\checkmark$ & $\times$ & $\times$ & $\times$ & $\times$ & \textbf{$\checkmark$} \\
eth       & $\checkmark$ & $\times$ & $\times$ & $\times$ & $\times$ & \textbf{$\checkmark$} \\ \hline\hline
AVE.      & \textbf{100\%} & 42.9\% & 35.7\% & 42.9\% & 50.0\% & \textbf{100\%} \\ \hline
\end{tabular}
}
\end{table}

\subsection{Benchmark and Baseline}
\noindent\textbf{Design Templates:}  
We construct a high-quality template library from the RTLRewriteark suite, curated with experienced Verilog engineers.
They are based on industrial design practices and over 50 scholarly works on RTL transformation.
Each template includes:
\begin{itemize}
    \item \textbf{RTL code} (\texttt{RTL}): optimized Verilog with favorable PPA (synthesized via \texttt{Synopsys Design Compiler} \cite{dc});
    \item \textbf{AST graph} ($\mathbb{G}$): JSON-serialized syntax tree capturing structural and logical semantics.
\end{itemize}
The library spans 55 templates across 21 RTL functions, including arithmetic datapaths, memory controllers, finite-state machines (FSMs), MUX-based control.

\noindent\textbf{Long Context Designs:}  
There are two types of large RTLs:
\begin{itemize}
\item \textbf{Single-Module RTL:}  
Flat, monolithic designs without module boundaries.
They pose major challenges for LLMs due to context overflow and entangled logic. 
\item \textbf{Multi-Module RTL:}  
Modular SoC-style designs with declared interfaces, yet still exceed context limits and contain non-trivial inter-module dependencies.
\end{itemize}
\Cref{tab:slong} and \Cref{tab:mlong} summarize their scale and PPA metrics, reflecting realistic industry-grade complexity.

\noindent\textbf{Baselines:}  
We compare against five strong baselines:
\begin{itemize}
\item \textbf{General LLMs:} GPT-4o and Gemini, representing state-of-the-art LLMs without RTL-specific tuning.
\item \textbf{RTL-Specific LLMs:} RTLCoder and RTLRewriter, both designed for Verilog generation and optimization.
\item \textbf{Compiler-Based Optimizers:} Yosys+Egg, an open-source complier, albeit inferior to \texttt{Synopsys Design Compiler} \cite{dc}.
\end{itemize}
All LLM-based methods use the same prompt format and context constraints for fair comparison.

\begin{table*}[t]
	\centering
	\caption{Comparisons of baseline approaches on single-module long context RTL benchmarks. ``RAT.'' represents the ratio of optimized RTL to original RTL on PPA metrics (Delay ($ps$), Area ($\mu m^2$), Power ($mW$)), as synthesized and reported by \texttt{Synopsys Design Compiler}. 
    Because Yosys \cite{wolf2013yosys} + Egg \cite{coward2022egg} performs no substantial logic optimization, the synthesized RTL shows no PPA improvement.}
	\label{tab:sper}
	\resizebox{1.0\linewidth}{!}
    {
\begin{tabular}{|c|ccc|ccc|ccc|ccc|ccc|ccc|}
\hline
\multirow{2}{*}{Des.} & \multicolumn{3}{c|}{Yosys \cite{wolf2013yosys} + Egg \cite{coward2022egg}} & \multicolumn{3}{c|}{GPT4o \cite{hurst2024gpt}} & \multicolumn{3}{c|}{Gemini \cite{Google2025Gemini}} & \multicolumn{3}{c|}{RTLCoder \cite{rtlcoder}} & \multicolumn{3}{c|}{RTLRewriter \cite{rtlrewriter}} & \multicolumn{3}{c|}{Ours} \\ \cline{2-19}
 & Delay & Area & Power & Delay & Area & Power & Delay & Area & Power & Delay & Area & Power & Delay & Area & Power & Delay & Area & Power \\ \hline \hline
add64     & 45.3 & 192.8 & 41.3 & / & / & / & 47.5 & 162.2 & 39.4 & / & / & / & 41.8 & 147.3 & 34.7 & \textbf{39.2} & \textbf{133.3} & \textbf{32.2} \\
mult32    & 669.7 & 259.0 & 220.1 & 547.2 & 194.6 & 164.7 & 552.8 & 198.0 & 172.5 & 567.6 & 216.2 & 183.2 & 499.7 & 178.0 & 149.7 & \textbf{463.9} & \textbf{158.8} & \textbf{145.9} \\
comput    & 1047.0 & 157.6 & 129.7 & / & / & / & / & / & / & / & / & / & / & / & / & \textbf{740.6} & \textbf{110.9} & \textbf{84.1} \\
traffic   & 10.2 & 66.7 & 1.6 & 8.5 & 53.3 & 1.3 & 9.0 & 53.3 & 1.3 & 8.8 & 56.0 & 1.3 & 7.8 & 47.7 & 1.2 & \textbf{7.4} & \textbf{44.7} & \textbf{1.1} \\
alu       & 1753.6 & 84.1 & 52.5 & 1413.1 & 66.6 & 43.2 & 1462.9 & 67.0 & 44.9 & 1450.7 & 74.0 & 46.6 & 1320.1 & 61.0 & 39.8 & \textbf{1122.9} & \textbf{56.1} & \textbf{39.6} \\
radix     & 285.0 & 33.6 & 7.4 & / & / & / & / & / & / & / & / & / & / & / & / & \textbf{189.2} & \textbf{22.6} & \textbf{5.0} \\
asyn      & 135.2 & 97.5 & 14.4 & 108.9 & 70.9 & 12.2 & / & / & / & 112.7 & 78.9 & 12.6 & 100.5 & 66.2 & 11.4 & \textbf{93.7} & \textbf{61.1} & \textbf{10.5} \\
accu      & 66.5 & 10.8 & 3.3 & 56.9 & 8.7 & 2.6 & 56.0 & 9.0 & 2.7 & 56.0 & 9.0 & 2.7 & 51.3 & 8.2 & 2.3 & \textbf{47.4} & \textbf{7.5} & \textbf{2.2} \\
fpu\_pre  & 44.6 & 70.4 & 51.4 & / & / & / & / & / & / & / & / & / & / & / & / & \textbf{31.1} & \textbf{42.6} & \textbf{33.4} \\
fpu\_post & 3174.5 & 149.1 & 115.0 & / & / & / & / & / & / & / & / & / & / & / & / & \textbf{2104.2} & \textbf{104.8} & \textbf{80.8} \\
mc\_sel   & 334.6 & 57.2 & 32.6 & / & / & / & / & / & / & / & / & / & / & / & / & \textbf{231.7} & \textbf{35.2} & \textbf{21.3} \\
mc\_rf    & 139.6 & 34.7 & 7.5 & 106.4 & 28.1 & 5.4 & / & / & / & 118.4 & 30.8 & 6.3 & 94.2 & 25.6 & 5.0 & \textbf{89.0} & \textbf{24.4} & \textbf{4.7} \\
buffer    & 58.5 & 117.6 & 27.7 & / & / & / & / & / & / & / & / & / & / & / & / & \textbf{42.5} & \textbf{76.7} & \textbf{17.1} \\
eth       & 60.7 & 96.6 & 13.5 & / & / & / & / & / & / & / & / & / & / & / & / & \textbf{40.5} & \textbf{65.3} & \textbf{10.1} \\ \hline\hline
AVE. & 558.9 & 102.0 & 51.3 & / & / & / & / & / & / & / & / & / & / & / & / & \textbf{373.1} & \textbf{67.2} & \textbf{35.1} \\ 
RAT. & 1.08 & 1.11 & 1.10 & 0.89 & 0.86 & 0.87 & 0.94 & 0.89 & 0.93 & 0.92 & 0.93 & 0.94 & 0.82 & 0.80 & 0.81 & \textbf{0.72} & \textbf{0.73} & \textbf{0.76} \\ \hline
\end{tabular}
}
\end{table*}

\begin{table*}[t]
	\centering
	\caption{Comparisons of baseline approaches on multi-module long context RTL benchmarks.}
	\label{tab:mper}
	\resizebox{1.0\linewidth}{!}{
\begin{tabular}{|c|ccc|ccc|ccc|ccc|ccc|ccc|}
\hline
\multirow{2}{*}{Des.} & \multicolumn{3}{c|}{Yosys \cite{wolf2013yosys} + Egg \cite{coward2022egg}} & \multicolumn{3}{c|}{GPT4o \cite{hurst2024gpt}} & \multicolumn{3}{c|}{Gemini \cite{Google2025Gemini}} & \multicolumn{3}{c|}{RTLCoder \cite{rtlcoder}} & \multicolumn{3}{c|}{RTLRewriter \cite{rtlrewriter}} & \multicolumn{3}{c|}{Ours} \\ \cline{2-19} & Delay & Area & Power & Delay & Area & Power & Delay & Area & Power & Delay & Area & Power & Delay & Area & Power & Delay & Area & Power \\ \hline \hline
ac97   & 107.5 & 1311.3 & 33.8  & 85.1  & 1124.8 & 27.7  & 80.5  & 1040.5 & 28.1  & 82.7  & 1123.4 & 28.1  & 77.9  & 1078.9 & 26.6  & \textbf{72.2}  & \textbf{1027.5} & \textbf{25.0}  \\
aes    & 122.3 & 999.0  & 16.0  & 97.3  & 881.4  & 13.4  & 93.7  & 819.7  & 13.5  & 97.9  & 885.4  & 13.8  & 88.0  & 847.4  & 12.8  & \textbf{82.5}  & \textbf{807.1}  & \textbf{12.3}  \\
crca   & 124.8 & 376.4  & 3.4   & 97.6  & 325.6  & 2.8   & 92.9  & 307.6  & 2.9   & 97.5  & 329.2  & 2.9   & 86.9  & 308.6  & 2.6   & \textbf{83.6}  & \textbf{299.6}  & \textbf{2.6}   \\
ecg    & 355.2 & 7435.7 & 46.8  & 286.9 & 6362.7 & 37.6  & 269.1 & 5988.9 & 38.9  & 276.9 & 6377.3 & 38.8  & 260.1 & 6055.0 & 36.7  & \textbf{246.7} & \textbf{5826.2} & \textbf{33.3}  \\ \hline \hline
AVE.   & 177.5 & 2530.6 & 25.0  & 141.7 & 2173.6 & 20.4  & 134.0 & 2039.2 & 20.8  & 138.8 & 2178.8 & 20.9  & 128.2 & 2072.5 & 19.7  & \textbf{121.3} & \textbf{1990.1} & \textbf{18.3}  \\ 
RAT.   & 1.17  & 1.12   & 1.13  & 0.93  & 0.96   & 0.92  & 0.88  & 0.90   & 0.94  & 0.92  & 0.96   & 0.94  & 0.85  & 0.91   & 0.88  & \textbf{0.80}  & \textbf{0.88}   & \textbf{0.82}  \\ \hline
\end{tabular}
}
\end{table*}

\subsection{Performance}
\noindent\textbf{Performance on Single-Module Long Context RTL.}  
This setting poses the core challenge for LLM-based RTL optimization: large monolithic designs (often $>$200 lines) without clear modular boundaries, causing severe context overflow and logic entanglement.
As shown in \Cref{tab:equ}, our method achieves 100\% functional equivalence across all test cases, thanks to AST-based partitioning guided by design templates. The reconstruction agent preserves signal consistency through logic-depth–aware ordering and graph-RAG prompting.
In contrast, prior methods succeed on less than 50\% of the cases, and completely fail when RTL line counts exceed 200—limiting their practical utility on real-world designs.
As shown in \Cref{tab:sper}, our method not only ensures full correctness but also achieves about 25\% average PPA improvement over original RTLs, surpassing all baselines in both quality and reliability.

\noindent\textbf{Performance on Multi-Module Long Context RTL.}  
We further validate our framework on SoC-style RTLs with explicit modules but long overall context and inter-module dependencies.
Here, our retrieval-then-optimize strategy remains effective: AST similarity–based retrieval yields structurally relevant exemplars, and MCTS-guided rewriting selects candidates with better PPA.
According to \Cref{tab:mper}, our method achieves up to 5\% PPA improvement over RTLRewriter, the strongest existing baseline, and about 15\% average improvement compared to original RTLs.

\begin{figure}[!tb]
	\centering
	 \hspace{-0.3in}
	\subfloat[On Single-Module RTLs.]
	{\centering
    \includegraphics[height=0.32\linewidth]{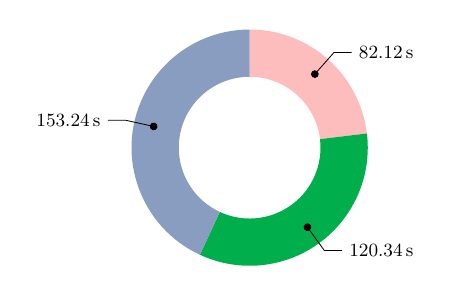} \label{fig:break1}
    }
	\subfloat[On Multi-Module RTLs.]
	{\centering
    \includegraphics[height=0.33\linewidth]{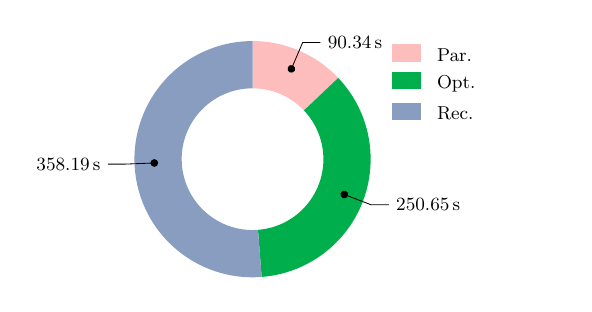} 
    \label{fig:break2} 
    }
	\caption{{{Average runtime breakdown of our work.}}} 
	\label{fig:runtimebreakdown}
\end{figure}

\noindent\textbf{Runtime Efficiency.}  
As shown in \Cref{fig:runtimebreakdown}, we report average runtime performance across both benchmarks.
The end-to-end average runtime is approximately 360s for single-module cases and 700s for multi-module ones. 
Notably, LLM inference is relatively efficient; the majority of the runtime overhead comes from logic synthesis (to evaluate PPA using \texttt{Synopsys Design Compiler}) \cite{dc} and simulation-based verification (to ensure functional equivalence using \texttt{ABC} \cite{brayton2010abc}).  
Our modular design enables parallel execution across submodules, allowing the framework to scale efficiently for industrial RTL use cases.

\begin{figure}[t]
	\centering
	\subfloat[]{\centering\includegraphics[height=0.42\linewidth]{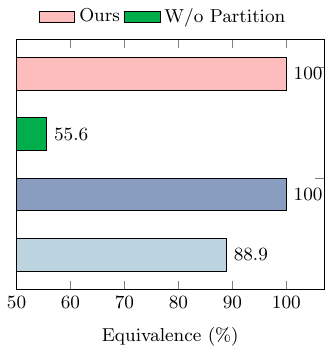} \label{fig:bar1}}
	\subfloat[]{\centering\includegraphics[height=0.42\linewidth]{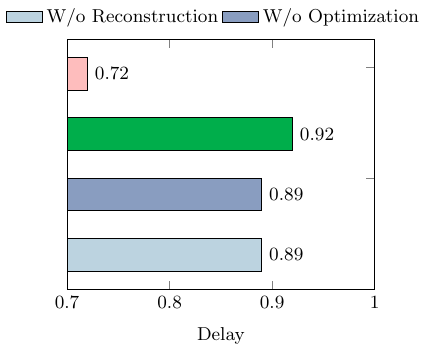} \label{fig:bar2} }\\
    \vspace{+0.1cm}
    \hspace{-0.45cm}
    	\subfloat[]{\centering\includegraphics[height=0.37\linewidth]{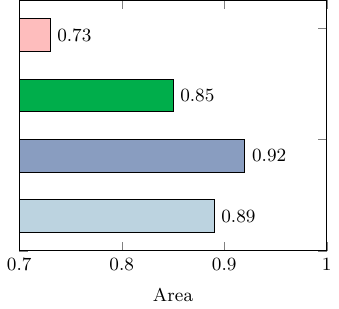} \label{fig:bar3}}
        \hspace{+0.35cm}
	\subfloat[]{\centering\includegraphics[height=0.37\linewidth]{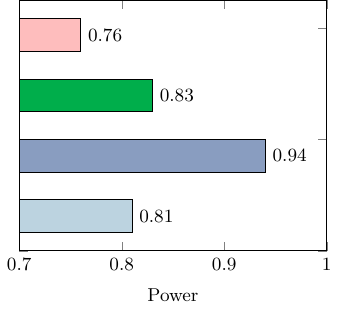} \label{fig:bar4} }
	\caption{{{Comparison among different schemes by (a) functional equivalence pass rate (\%), (b) delay, (c) area, and (d) power. All these values are normalized by original RTL design results reported via \texttt{Synopsys Design Complier}.}}} 
	\label{fig:bar}
\end{figure}
\subsection{Ablation Studies}
To evaluate each component's contribution, we conduct ablations on three aspects: (1) graph similarity-driven partitioning, (2) multimoal retrieval-augmented optimization, and (3) logic-aware reconstruction. 
As shown in \Cref{fig:bar}, results are measured by functional equivalence pass rate and normalized PPA metrics.

\noindent\textbf{(1) W/o Partition: Random Subtree Splitting.}  
Instead of using Tree-DP with template-guided graph similarity, we randomly split the long-context AST $\mathbb{G}_{\text{long}}$ into fixed-size subgraphs. Despite using the same optimization agent, the functional pass rate drops by 44.4\%, and delay/area increase due to poorly defined module boundaries.

\noindent\textbf{(2) W/o Optimization: No Multimodal Retrieval Guidance.}  
We remove RTL+AST-based retrieval and directly prompt the LLM with common RTL optimization instructions like RTLRewriter \cite{rtlrewriter}. Lacking structural grounding, the generated RTL becomes generic and misaligned, leading to noticeable PPA degradation.

\noindent\textbf{(3) W/o Reconstruction: Naive Concatenation.}  
We skip logic-aware ordering and graph prompting during reassembly, instead concatenating submodules. 
This results in broken control/data paths and I/O mismatches, reducing functional correctness by 11.1\% and impacting downstream PPA.

\section{Conclusion}
\label{sec:conclu}
This work presents a complete framework for long-context RTL optimization, comprising three collaborative agents:  
(1) a graph similarity–guided partition agent that decomposes large ASTs into semantically meaningful subtrees;  
(2) a multimodal optimization agent that jointly leverages AST structure and RTL semantics for retrieval-augmented generation with MCTS-based candidate selection;  
(3) a logic-aware reconstruction agent that ensures functional equivalence via graph-RAG prompting and logic-aware reassembly.  
Experiments on challenging RTL designs confirm 100\% functional equivalence and consistent PPA improvements over state-of-the-art baselines, demonstrating the framework’s scalability and real-world applicability.

\clearpage
\bibliographystyle{IEEEtran}
\bibliography{Top-sim,RTL}

\end{document}